\documentclass[11pt, oneside]{iopart}
\pdfoutput=1

\usepackage{iopams}
\expandafter\let\csname equation*\endcsname\relax
\expandafter\let\csname endequation*\endcsname\relax
\usepackage{graphicx,float,cite}
\usepackage{amsmath,amssymb}
\usepackage{tikz}
\usepackage[utf8]{inputenc}

\usepackage{xcolor}
\usepackage[normalem]{ulem}

\usepackage[colorlinks,bookmarks=false,citecolor=orange,linkcolor=blue,hyperfootnotes=true,urlcolor=blue]{hyperref}

\begin{document}

\title[Lack of thermalization in (1+1)-d QCD at large $N_c$]{\boldmath Lack of thermalization in (1+1)-d QCD at large $N_c$}

\author{Axel Cort\'{e}s Cubero$^{\ast,\dagger}$ and Neil J Robinson$^\dagger$}
\address{$^\ast$Institute for Theoretical Physics, Center for Extreme Matter and Emergent Phenomena, Utrecht University, Princetonplein 5, 3584 CC Utrecht, The Netherlands\\ \vspace{1mm}
  $^\dagger$Institute for Theoretical Physics, University of Amsterdam, Postbus 94485, 1090~GL Amsterdam, The Netherlands}

\begin{abstract}
  Motivated by recent works aimed at understanding the status of equilibration and the eigenstate thermalization hypothesis in theories with confinement, we return to the 't~Hooft model, the large-$N_c$ limit of (1+1)-d quantum chromodynamics. This limit has been studied extensively since its inception in the mid-1970s, with various exact results being known, such as the quark and meson propagators, the quark-antiquark interaction vertex, and the meson decay amplitude. We then argue this model is an ideal laboratory to study non-equilibrium phenomena, since it is manifestly non-integrable, yet one retains a high level of analytic control through large-$N_c$ diagrammatics. We first elucidate what are the non-equilibrium manifestations of the phenomenon of large-$N_c$ volume independence. We then find that within the confined phase, there is a class of initial states that lead to a violation of the eigenstate thermalization hypothesis, i.e. the system never thermalizes. This is due to the existence of heavy mesons with an extensive amount of energy, a phenomenon that has been numerically observed recently in the quantum Ising chain.

  $~$

  \noindent{\bfseries Keywords.} Thermalization, Eigenstate thermalization hypothesis, large-$N_c$ volume independence, 't~Hooft model 
\end{abstract}

\maketitle
\flushbottom

\section{Introduction}

Interest in non-equilibrium quantum many-body systems has undergone a resurgence in the past decade, both from the point of view of experiment~\cite{langen2015ultracold} and theory (see, e.g.,~\cite{calabrese2016introduction} and the accompanying review articles). The condensed matter physics community has been motivated by advances in the field of cold atomic gases that rendered the non-equilibrium dynamics of isolated quantum systems observable~\cite{kinoshita2006quantum,polkovnikov2011colloquium,langen2015ultracold}. At the same time, the high-energy physics community was spurred on by experimental studies of the quark-gluon plasma, first at the Relativistic Heavy Ion Collider, and later at the Large Hadron Collider~\cite{shuryak2005what,blaizot2012boseeinstein,pasechnik2017phenomenological,shuryak2017strongly}. In both communities it was rather rapidly realized that a number of fundamental issues were not as well understood as one might have thought, including the questions of how, when and why isolated quantum systems thermalize. A number of seminal works addressed these questions (see~\cite{dalessio2016from} and references therein), elucidating the central role conservation laws play in restricting non-equilibrium dynamics. Indeed, such restrictions may be sufficient to avoid thermalization entirely~\cite{rigol2007relaxation}.

Rapidly a dichotomy emerged in describing thermalization in quantum many-body systems~\cite{rigol2007relaxation,rigol2008thermalization}. On the one hand lay systems with extensively many local conservation laws, such as quantum integrable~\cite{korepin1993quantum} and many-body localized~\cite{nandkishore2015manybody,alet2018manybody,parameswaran2018manybody} systems. Here the constraints imposed by the conservation laws restrict the dynamics to such an extent that thermalization is avoided~\cite{rigol2007relaxation}. Instead, after long times local expectation values in such non-equilibrium systems equilibrate and are described in terms of a generalized Gibbs ensemble~\cite{rigol2007relaxation,vidmar2016generalized,ilievski2016quasilocal}, a generalization of the thermal ensemble to the case of many conserved quantities. Such systems retain an extensive amount of information about the initial state, thanks to the many quantities conserved under the dynamics. On the other hand are generic systems, possessing just a small (intensive) number of conserved quantities, where information about the initial state is scrambled and lost under non-equilibrium dynamics. In the long time limit, measurements in the non-equilibrium system equilibrate to those of the thermal ensemble, whose temperature is fixed by the energy of the initial state~\cite{rigol2008thermalization}. The description of non-equilibrium dynamics of generic systems is extremely challenging. Ideas that hark back to the early 1990s~---~the eigenstate thermalization hypothesis~---~play a crucial role in our understanding of thermalization in isolated quantum systems~\cite{deutsch1991quantum,srednicki1994chaos,dalessio2016from}. In section~\ref{sec:eth}, we will discuss further the eigenstate thermalization hypothesis, and the cases in which it is violated. 

This integrable/generic dichotomy has been extremely successful in describing equilibration and thermalization in isolated quantum many-body  systems~\cite{rigol2008thermalization,rigol2007relaxation,vidmar2016generalized,ilievski2016quasilocal}. However over the past couple of years it has become increasingly apparent that there are simple quantum many-body systems that fall outside this dichotomy (see, e.g., refs.~\cite{turner2017quantum,turner2018quantum,ho2019periodic,khemani2019signatures,lin2019exact,choi2018emergent,iadecola2019quantum,bull2019systematic,moudgalya2018exact,moudgalya2018entanglement,sala2019ergodicitybreaking,khemani2019local,craps2014gravitational,craps2015holographic,dasilva2016holographic,myers2017holographic,kormos2016realtime,rakovszky2016hamiltonian,hodsagi2018quench,james2019nonthermal,robinson2019signatures,buyens2014matrix,buyens2016confinement,buyens2017finiterepresentation,buyens2017realtime,banuls2017density,akhtar2018symmetry,park2019glassy,mazza2018suppression,lerose2019quasilocalized,schecter2019weak,moudgalya2019quantum}). These systems are generic, in the sense that they do not possess extensively many conserved quantities (i.e. they are non-integrable, as can be seen via level spacing statistics), but none-the-less are not compatible with the eigenstate thermalization hypothesis. As a result, these systems show anomalous non-equilibrium dynamics, such as long-lived persistent oscillations in the time-evolution of local observables and an absence of thermalization. On the experimental side, such anomalous non-equilibrium dynamics was recently observed in a (1+1)-d lattice system of ultracold Rydberg atoms~\cite{bernien2017probing}. The subset of these non-integrable systems that violate the eigenstate thermalization hypothesis is discussed in detail in section~\ref{ethconfinement}. 

The dynamics of thermalization, and the eigenstate thermalization hypothesis, are non-perturbative phenomena in the interaction strength. Integrable systems are known for being exactly solvable, and there are many powerful analytical tools that can be applied to the study of non-equilibrium dynamics (see, e.g., refs.~\cite{essler2016quench,caux2016quench}). When studying the non-equilibrium properties of generic, non-integrable systems, however, one necessarily has less analytic control and must rely on numerical methods. We are then faced with a fundamental problem: How to study strictly non-perturbative phenomena in a (necessarily) non-integrable system? Here, we will overcome this difficulty by applying tools from the large-$N_c$ expansion of low dimensional gauge theories (see ref.~\cite{lucini2013sun} for a review of such approaches).

\subsection{The eigenstate thermalization hypothesis}
\label{sec:eth}

The eigenstate thermalization hypothesis~\cite{deutsch1991quantum,srednicki1994chaos} lies at the heart of understanding thermalization in isolated quantum many-body systems, so it is worth recapitulating some essential details. Further information, and applications of the eigenstate thermalization hypothesis, can be found in the recent review articles~\cite{reimann2015eigenstate,deutsch2018eigenstate}.

The eigenstate thermalization hypothesis aims to address the question of how and why quantum systems thermalize. Let us consider the explicit example of a pure state $|\Psi\rangle$ time-evolving according to a local Hamiltonian $H$. Local observables $O$ in the time-evolved state are given by
\begin{align}
  \langle O(t) \rangle \equiv \langle \Psi(t) | O |\Psi(t)\rangle  = \sum_{i,j} c_i c^\ast_j e^{i(E_j-E_i)t} \langle E_j | O | E_i\rangle.
\end{align}
Here we denote the eigenstates of the Hamiltonian with energy $E_i$ as $|E_i\rangle$, and $c_i = \langle E_i |\Psi\rangle$ describes the projection of the state $|\Psi\rangle$ onto the eigenstate $|E_i\rangle$. For a generic case with no degeneracies, $E_i \neq E_j$ $\forall i\neq j$,
\begin{align}
  \langle O(t) \rangle = \sum_i |c_i|^2 \langle E_i | O |E_i \rangle + \sum_{i\neq j} c_i c_j^\ast e^{i(E_i-E_j)t} \langle E_j | O | E_i \rangle. \label{de}
\end{align}
The second term on the right hand side is oscillatory. Thus in the long time limit, under a stationary phase evaluation of the sum (or taking the time-average), it averages to zero and so 
\begin{align}
  \lim_{t\to\infty} \langle O(t) \rangle = \sum_i |c_i|^2 \langle E_i | O |E_i \rangle.  \label{longtime}
\end{align}
This is known as the diagonal ensemble result for the long-time limit, as it features only diagonal matrix elements of the operator $O$ in the eigenbasis. It would seem that eq.~\eqref{longtime} depends sensitively on the initial state $|\Psi\rangle$ through the superposition coefficients $c_i$, in the sense that small (local) changes in $|\Psi\rangle$ can significantly change the superposition coefficients $c_i = \langle E_i | \Psi\rangle$. Yet if the system is thermalizing, we expect eq.~\eqref{de} to coincide with the microcanonical ensemble average result
\begin{align}
  \langle O \rangle_{\text{MCE}} = \frac{1}{N_{\Delta E}}\sum_{\substack{i:\\|E_i - E_\Psi|<\Delta E}} \langle E_i | O | E_i \rangle. \label{mce}
\end{align}
Here  $E_\Psi = \langle \Psi | H | \Psi\rangle$ is the energy of the state $|\Psi\rangle$, $\Delta E$ is a small energy window for the microcanonical averaging, and $N_{\Delta E}$ is the number of states within this energy window. Why and how can this be the case? 

The eigenstate thermalization hypothesis gives a set of conditions under which eq.~\eqref{longtime} coincides with eq.~\eqref{mce} (when energy $E_\Psi$ is extensive). For eigenstates $|E_i\rangle$ and $|E_j\rangle$ within a given symmetry sector, matrix elements of the local observables $O$ should satisfy (see, e.g., ref.~\cite{mondaini2017eigenstate})
\begin{equation}
  \langle E_i | O | E_j \rangle = f_O(E)\delta_{i,j} + e^{-S(E)/2}g_O(E,\omega) R_{ij}, \label{eth}
\end{equation}
with $E$ being the average energy of the eigenstates, i.e. $E=(E_i+E_j)/2$, and $\omega$ being the energy difference, $\omega = E_i - E_j$. Here $f_O(E)$ is a smooth function of the energy $E$ for a given operator $O$. Off-diagonal elements are suppressed by the thermodynamic entropy $S(E)$ and related to a smooth function $g_O(E,\omega)$ of the average energy and the energy difference, as well as a random variable $R_{ij}$ with zero mean and unit variance.

Let us focus on the diagonal part of eq.~\eqref{eth}, which conjectures the form of expectation values of a local operator within a given state
\begin{equation}
  \langle E | O | E \rangle = f_O(E). \label{ethdiag}
\end{equation}
That is, for a given operator $O$  expectation values within an eigenstate are a smooth function of the energy $E$ alone. In such a scenario, the microcanonical ensemble can be truncated to a single state ($\Delta E = 0,\,N_{\Delta E}=1$) in eq.~\eqref{mce}. Such a system is then thermalizing, by construction. It is worth mentioning that expectation values in the finite volume (rather than in the thermodynamic limit) can have some finite variance from eq.~\eqref{ethdiag}, which should shrink to zero in the infinite volume limit. Studies that examine eq.~\eqref{ethdiag} are, by necessity, usually undertaken via exact diagonalization of small systems --- see the review article~\cite{deutsch2018eigenstate} and references therein.

\subsubsection{Systems that violate the eigenstate thermalization hypothesis} 
It so far seems to be the case that thermalizing isolated quantum many-body systems satisfy eq.~\eqref{ethdiag}, see~\cite{deutsch2018eigenstate}. There are, however, a number of examples of quantum many-body systems where the eigenstate thermalization hypothesis is violated. In the next section, section~\ref{ethconfinement}, we discuss in more detail a particular class of such systems of relevance to this work, here we will briefly mention some other examples. 

The most-well studied cases of violations of the eigenstate thermalization hypothesis are those of integrable quantum systems. Following the quantum Newton's cradle experiment~\cite{kinoshita2006quantum}, it was rapidly realized that such systems do not thermalize due to the presence of extensively many conservation laws~\cite{essler2016quench,vidmar2016generalized,ilievski2016quasilocal}. In eigenstate thermalization hypothesis studies, the presence of many (hidden) conservation laws is revealed by a clear violation of eq.~\eqref{ethdiag}; at fixed energy $E$ there exist eigenstates with very different expectation values of local observables, and the variance of these remains finite in the thermodynamic limit. Many-body localized systems, in which there is an emergent integrability, also violate the eigenstate thermalization hypothesis is a similar manner~\cite{nandkishore2015manybody,alet2018manybody,parameswaran2018manybody}.

Recently, it has been realized that there exist non-integrable quantum systems that also violated the eigenstate thermalization hypothesis. Perhaps the best studied case of this so far is the PXP model, in which there are a polynomial-in-the-system-size number of eigenstate thermalization hypothesis violating eigenstates, often called many-body quantum scars~\cite{turner2017quantum,turner2018quantum,ho2019periodic,khemani2019signatures,lin2019exact,choi2018emergent,iadecola2019quantum}. Other examples include certain spin models, such as generalized Affleck-Kennedy-Lieb-Tasaki chains~\cite{moudgalya2018exact,moudgalya2018entanglement} and spin-1 XY models~\cite{schecter2019weak}, as well as fracton models~\cite{sala2019ergodicitybreaking,khemani2019local} and the thin torus limit of Landau levels at certain fillings~\cite{moudgalya2019quantum}.

As we have already mentioned, the non-perturbative nature of these phenomena combined with the non-integrable character of these systems restricts these studies to being either purely numerical or based upon ill-controlled analytical approximations. In this work we are presented with a rare opportunity, through the use of the large-$N_c$ expansion for low dimensional gauge theories~\cite{lucini2013sun}, to study the breaking of the eigenstate thermalization hypothesis analytically.  

\subsection{Non-equilibrium dynamics and the eigenstate thermalization hypothesis in models with confinement}
\label{ethconfinement}
There is yet another class of non-integrable models, which would generally be thought to be thermalizing~\cite{rigol2008thermalization}, that display anomalous non-equilibrium behavior. Models with confinement can exhibit an absence of thermalization, unusual non-equilibrium dynamics, and a violation of the eigenstate thermalization hypothesis. Such behavior has been seen in both continuum and lattice models, with wide-ranging examples including: holographic models~\cite{craps2014gravitational,craps2015holographic,dasilva2016holographic,myers2017holographic}, certain perturbed conformal field theories~\cite{kormos2016realtime,rakovszky2016hamiltonian,hodsagi2018quench,james2019nonthermal,robinson2019signatures}, the anisotropic 2D Ising field theory~\cite{james2019nonthermal}, the Schwinger model~\cite{buyens2014matrix,buyens2016confinement,buyens2017finiterepresentation,buyens2017realtime,banuls2017density,akhtar2018symmetry}, the extended Bose-Hubbard model~\cite{park2019glassy}, and various spin chains~\cite{kormos2016realtime,james2019nonthermal,mazza2018suppression,lerose2019quasilocalized}. Our aim in this work will be to establish similar anomalous dynamics and an absence of thermalization in 't~Hooft's model of (1+1)-d quantum chromodynamics (QCD). Before continuing to our study, we will first spend some time reminding the reader of the anomalous non-equilibrium behavior exhibited by models with confinement.

In holographic models quenches within the confined phase (i.e. those quenches where the energy injected is not sufficient to drive a confinement-deconfinement transition) have been considered in refs.~\cite{craps2014gravitational,craps2015holographic,dasilva2016holographic,myers2017holographic}. In the holographic hard wall model, for example, on the gravity side it was found there is insufficient energy to form a black brane (as expected to occur for thermalization) and instead long-lived oscillatory ``scattering solutions'' govern the physics~\cite{craps2014gravitational}. Similar phenomenology was observed in the AdS soliton model~\cite{craps2015holographic}. Later works examined a number of thermalization indicators, including those not reliant upon the formation of a black brane or black hole horizon, finding no signatures of thermalization~\cite{myers2017holographic}. Both local and non-local observables were observed to undergo anomalous long-lived oscillatory dynamics in these holographic confined models~\cite{myers2017holographic}.  We do note, however, that these behaviors may be related to the strict $N_c=\infty$ limit taken within these works, where some remnants of integrability are present. (In the $N_c=\infty$ limit, heavy mesons have infinite lifetime in the considered holographic model, thus there are many constants of motion in this limit. Furthermore, the quench considered in Ref.~\cite{myers2017holographic} generates only single meson states.)

The unusual physics exhibited by holographic confined models has also been realized in a very different setting: one-dimensional quantum spin chains. The elementary excitations of such systems, spinons, can be confined by the presence of a magnetic field or long-range exchange interactions. This observation goes back to the seminal work of McCoy and Wu on the two-dimensional Ising field theory and their study of the poles of the two-particle $S$-matrix~\cite{mccoy1978twodimensional}. Two-spinon confined excitations, analogous to the mesons of QCD, have been observed in beautiful inelastic neutron scattering~\cite{coldea2010quantum} and THz spectroscopy~\cite{morris2014hierarchy} experiments on the ferromagnet CoNb$_2$O$_6$. Identical physics can also arise in antiferromagnets when the longitudinal magnetic field is staggered (oscillating in sign from site-to-site in the chain), as illustrated by SrCo$_2$V$_2$O$_8$~\cite{wang2015spinon,wang2016from,bera2017spinon}. Theoretical studies of both lattice and continuum descriptions of such spin chains have observed anomalous non-equilibrium dynamics: the suppression of light-cone spreading of correlations~\cite{kormos2016realtime,james2019nonthermal,mazza2018suppression,lerose2019quasilocalized}, long-lived oscillations in local observables~\cite{rakovszky2016hamiltonian,hodsagi2018quench,james2019nonthermal,robinson2019signatures}, and an absence of thermalization~\cite{james2019nonthermal,robinson2019signatures}. Importantly, these anomalies can be linked to an explicit violation of the ETH, with single meson eigenstates being distinctly non-thermal\footnote{In the sense that expectation values of local operators in single meson eigenstates do not agree with the microcanonical average (thermal) result.} and extending far into the many-body spectrum~\cite{james2019nonthermal,robinson2019signatures}.  

\subsection{Overview of our main results}

With this discussion of the unusual non-equilibrium physics of confined models in mind, in the remainder of this work we turn our attention to (1+1)-d QCD (known as the 't~Hooft model in the large-$N_c$ limit). The physical spectrum of the 't~Hooft model is qualitatively similar to the confining Ising model, consisting of a tower of meson bound states (see recent eigenstate thermalization studies~\cite{james2019nonthermal,robinson2019signatures} and earlier studies of the spectrum~\cite{fonseca2006ising}). As we will show, this implies that some of the non-equilibrium aspects of the two models are very similar. Despite their similarities, the large-$N_c$ expansion gives us significantly more analytic control over the 't~Hooft model (for the Ising model only numerical non-equilibrium studies have been possible~\cite{kormos2016realtime,rakovszky2016hamiltonian,hodsagi2018quench,james2019nonthermal,robinson2019signatures}).

Taking the number of colors, $N_c$ to be large (as first introduced by 't~Hooft~\cite{thooft1974twodimensional}), a useful correspondence arises between powers of $N_c$ and number of particles in matrix elements of gauge-invariant operators. At thermal equilibrium, we use this to show that thermal corrections to expectation values of gauge-invariant operators are suppressed by higher powers of $1/N_c$. This result is related to the well-known phenomenon of large-$N_c$ volume independence~\cite{eguchi1982reduction,kovtun2007volume}. We show how this concept can be generalized to non-equilibrium physics, leading to a hierarchy of $1/N_c$-suppressed terms.
 
We then use the $1/N_c$ expansion, in combination with properties of the matrix elements that appear within this expansion, to argue that quenches from finite energy density states containing few heavy, high-energy mesons do not thermalize (in the sense that the long-time limit of local observables does not approach the relevant thermal value). This statement holds even once the non-perturbative corrections that can arise in the long-time limit are taken into account. At the heart of our argument is the absence of annihilation poles in one-particle-to-one-particle matrix elements, which can be traced back to the local nature of meson excitations with respect to local observables.

We discuss our results in the context of the non-thermal states already observed in quantum magnets with confinement~\footnote{Note that such systems can be related, via an infrared correspondence, with compactified QCD~\cite{sulejmanpasic2017confinement}.}. The absence of thermalization implies a violation of the eigenstate thermalization hypothesis. Our results fits into an emerging picture in which theories with confinement can exhibit unusual thermalization properties.

\subsection{Layout}
The remainder of this work proceeds as follows. In section~\ref{sec:quarkint} we remind the reader of various exact results in the 't~Hooft model in the large $N_c$ limit. Section~\ref{sec:mesonints} discusses the interactions between meson excitations of the model, while in section~\ref{sec:correspondence} we discuss a useful particle number/powers of $N_c$ correspondence. We use this correspondence in section~\ref{sec:volumeindependence} to find a non-equilibrium generalization of the concept of large-$N_c$ volume independence, and in section~\ref{sec:quenches} to discuss the non-equilibrium dynamics following a quantum quench, which leads us to describe mesons as non-thermal states in section~\ref{sec:nonthermal}. We conclude in section~\ref{sec:conclusions}.

\section{Exact results in the 't~Hooft model}
\label{sec:exact}

\subsection{Planar Feynman diagrams, quark propagator and interaction vertex}
\label{sec:quarkint}

We consider herein the theory of QCD in (1+1)-d described by the Lagrangian,
\begin{equation}
\mathcal{L}=-\frac{1}{4}G_{\mu\nu\,i}^{j}G^{\mu\nu\,i}_{j}+\bar{q}^i\left({\rm i}\gamma^\mu D_\mu-m\right)q_i,\label{lagrangian}
\end{equation}
where the indices $i,j=1,\dots,N_c$ run over different colors of quarks and gluons, $q_i$ is a fermionic quark field, which takes vector values in $SU(N_c)$ color space, and there is a $SU(N_c)$ matrix-valued (in the adjoint representation of $SU(N_c)$) gauge field $A_{i\,\mu}^{j}$,  which is included in the field strength and covariant derivative, respectively defined as
\begin{align}
G_{\mu\nu\,i}^{j}\equiv&\partial_\mu A_{i\,\nu}^j-\partial_\nu A_{i\,\mu}^j+g\left[A_\mu,A_\nu\right]_{i}^j,\nonumber\\
D_\mu q_i\equiv&\partial_\mu q_i+gA_{i\,\mu}^j\,q_j.
\end{align}
The parameters in~\eqref{lagrangian} are the quark mass, $m$, and the strong coupling constant, $g$. 

It is particularly convenient to study this model within the light-cone gauge, where one of the components of the gauge field is fixed to zero,
\begin{equation}
A_-\equiv\frac{1}{\sqrt{2}}\big(A_0-A_1\big)=0,
\end{equation}
corresponding to the light-cone coordinates $x^{\pm}=x_{\mp}=\frac{1}{\sqrt{2}}(x^0-x^1)$. In this gauge the Lagrangian is given by the simple form
\begin{equation}
  \mathcal{L}=-\frac{1}{2}{\rm Tr}\big(\partial_- A_+\big)^2
  -\bar{q}\big(\gamma^\mu\partial_\mu +m+g\gamma_- A_+\big)q. \label{gaugeFixedL}
\end{equation}
This gauge is particularly convenient because there are no Faddeev-Popov ghosts and it is easy to see that there is no vertex for gluon self interactions.

It was shown by 't~Hooft that in the $N_c\to\infty$ limit, while keeping $g^2N_c$ fixed, some aspects of this model become exactly solvable. We will refer to the $N_c\to\infty$ limit of the model defined in~\eqref{lagrangian} as the 't~Hooft model. Moreover, while the 't~Hooft model becomes analytically tractable, the dynamics are sufficiently rich that the model is still qualitatively similar to the finite $N_c$ case. As we will see, it can be shown that the quarks in the 't~Hooft model are confined, and there is an infinite tower of meson (quark-antiquark) bound states.

\begin{figure}
  \begin{center}
    \includegraphics[width=0.8\textwidth,trim = 0 50 20 60,clip]{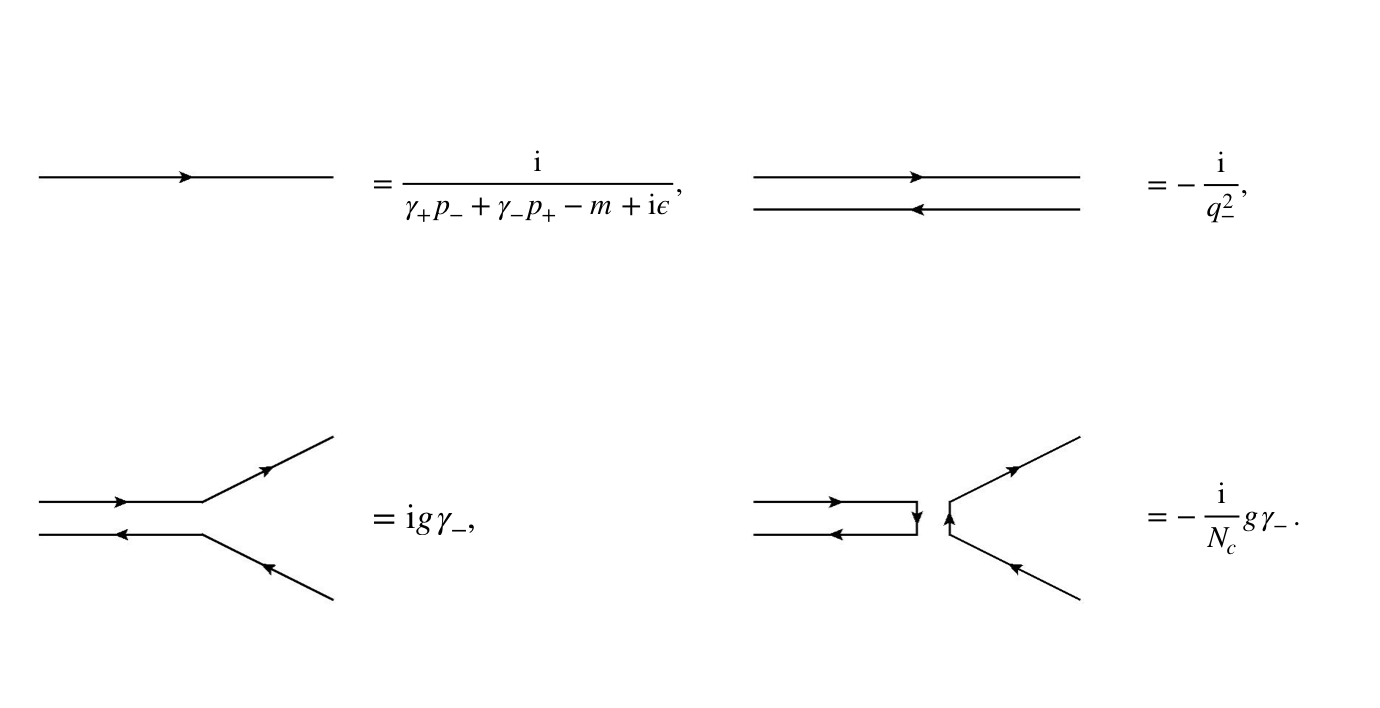}
  \end{center}
  \caption{The Feynman rules for~\eqref{lagrangian} describing (1+1)-d QCD with $N_c$ colors.}
  \label{fig:feynmanRules}
\end{figure}

The large-$N_c$ limit is simplified by the fact that the leading Feynman diagrams in orders of $1/N_c$ have a planar structure~\cite{thooft1974planar}. Diagrammatically,  quarks, which carry one fundamental color index, can be represented as a single line, while gluons, in the adjoint representation with two color indices, can be drawn as a double line, as pictured in figure~\eqref{fig:feynmanRules}.
In the large-$N_c$ limit, the leading diagrams are those which can be drawn on a single plain, while the crossing of color lines is suppressed.

\begin{figure}
  \begin{center}
    \includegraphics[width=0.7\textwidth,trim=0 0 0 0,clip]{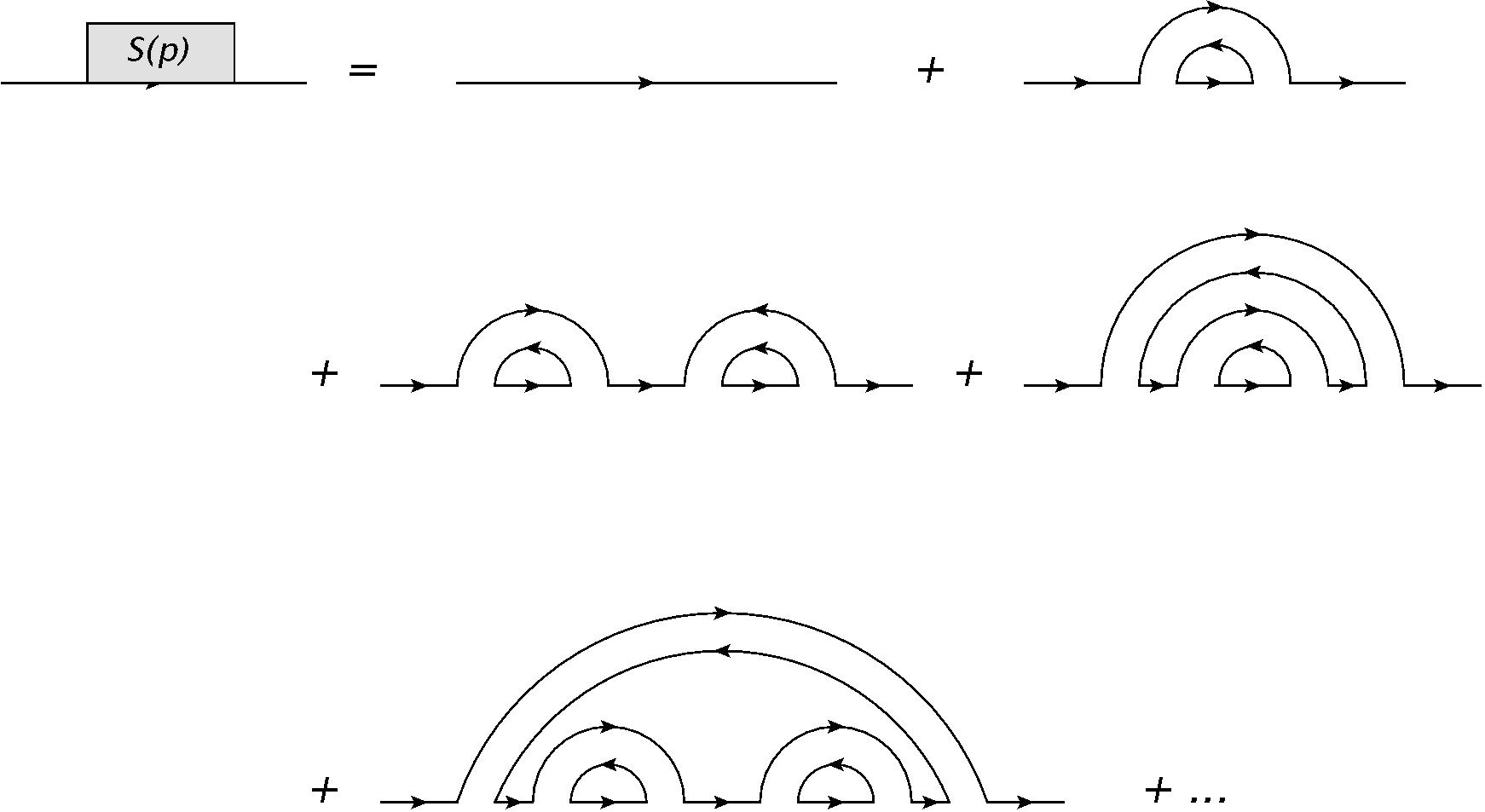}
  \end{center}
  \caption{In the large-$N_c$ limit, corrections to the fully-dressed quark propagator (represented herein as a rectangular blob) in the 't~Hooft model come in the form of rainbow-like diagrams, illustrated here. }
  \label{fig:rainbow}
\end{figure}

In the light cone gauge, the fact that there are no gluon self-interactions [cf.~\eqref{gaugeFixedL}], combined with the simplification of considering only planar diagrams, means that the quark propagator can be computed exactly, up to $1/N_c$ corrections. This computation can be easily understood diagrammatically, as was shown by 't~Hooft \cite{thooft1974twodimensional}. The large-$N_c$ limit implies that the only corrections to the quark propagator come from rainbow-like diagrams, involving emission and absorption of gluon lines that do not intersect each other, as shown in figure~\ref{fig:rainbow}. 

The full dressed propagator, $S(p)$ can be written in terms of the bare propagator, $S_0(p)$ as
\begin{equation}
  S(p) = \frac{S_0(p)}{1+{\rm i}\Sigma(p) S_0(p)}
  =\frac{{\rm i}p_-}{2p_+p_--m^2-p_-\Sigma(p)+{\rm i}\epsilon},
\end{equation}
where $\Sigma(p)$ is the sum of one-particle-irreducible diagrams.

\begin{figure}
  \begin{center}
      \includegraphics[width=0.8\textwidth]{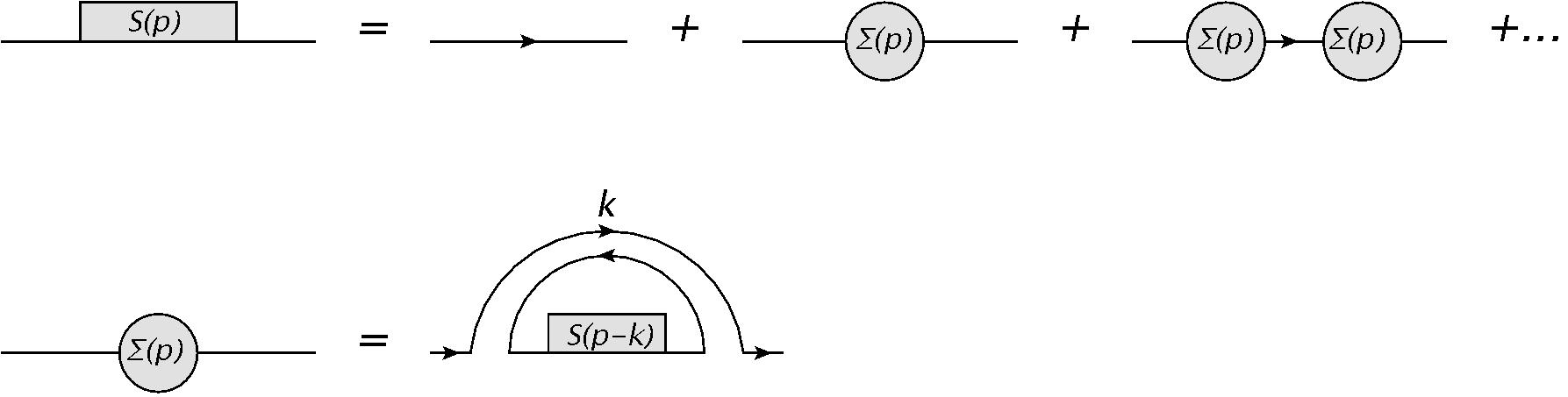}
  \end{center}
  \caption{The consistency conditions for the full dressed quark propagator $S(p)$, in terms of the one-particle-irreducible contributions, $\Sigma(p)$, from which one obtains eq.~\eqref{selfconsistentsigma}.}
  \label{fig:propagator}
\end{figure}

The full dressed propagator $S(p)$ can be computed exactly as the solution of an integral equation that follows from the consistency condition pictured in figure~\ref{fig:propagator}.
This gives
\begin{equation}
-{\rm i}\Sigma(p)=\int\frac{\mathrm{d}k_-\mathrm{d}k_+}{(2\pi)^2}\frac{{\rm i}}{k^2}S(p-k)(-2{\rm i}g)^2N_c,\label{selfconsistentsigma}
\end{equation}
where the integral is infrared divergent, so it is necessary to introduce a low momentum cutoff $\lambda$. It can be shown~\cite{thooft1974twodimensional} that the solution of~\eqref{selfconsistentsigma} is 
\begin{equation}
\Sigma(p)=-\frac{g^2 N_c}{\pi}\left(\frac{{\rm sgn(p_-)}}{\lambda}-\frac{1}{p_-}\right),
\end{equation}
and the full dressed propagator is
\begin{equation}
S(p)=\frac{-{\rm i}p_-}{m^2-g^2N_c/\pi+2p_+p_-+g^2N_c\vert p_-\vert/\pi\lambda-{\rm i}\epsilon}.
\label{propagator}
\end{equation}
Due to the presence of the cutoff $\lambda$ in the propagator, which has to be taken to zero, the pole in the propagator shifts to $p_+\to\infty$. This can be interpreted as the fact that the effective mass of a single quark diverges, i.e. quarks are confined and can only exist as bound states.

With knowledge of the full quark propagator at hand, it is then possible to study the full quark-quark interaction vertex, $T_{\alpha \beta,\gamma\delta}(p,p^\prime;r)$. 
\begin{figure}
  \begin{center}
	\includegraphics[width=0.9\textwidth]{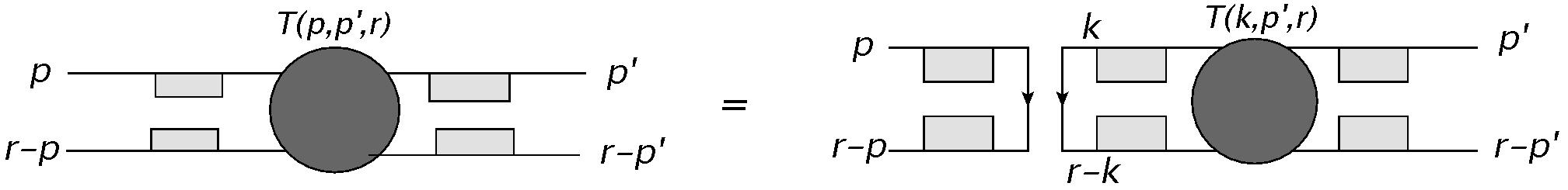}
  \end{center}
  \caption{Bethe-Salpeter consistency condition for the quark-quark vertex (represented herein as a dark-gray circular blob). As above, the gray rectangular boxes represent full quark propagators.}
  \label{fig:vertex}
\end{figure}
It was shown in~\cite{callan1976twodimensional,einhorn1976confinement} that this vertex satisfies the consistency condition (pictured in figure~\ref{fig:vertex}),
\begin{align}
  T_{\alpha\beta,\gamma\beta}(p,p^\prime,r)=&
  \frac{{\rm i}g^2}{(p_--p_-^\prime)^2}(\gamma_-)_{\alpha\gamma}(\gamma_-)_{\beta\delta}\nonumber\\
  & +{\rm i}g^2N_c\int\frac{\mathrm{d}^2k}{(2\pi)^2}
  \frac{(\gamma_-)_{\alpha\epsilon}(\gamma_-)_{\beta\lambda}}{(k_--p_-)^2}
  S(k)_{\epsilon\mu}S(k-r)_{\lambda\nu}T_{\mu\nu;\gamma\delta}(k,p^\prime;r).
\label{vertex}
\end{align}
This can be simplified by introducing the function
\begin{equation}
\phi(p_-,p^\prime_-;r)=\int \mathrm{d}p_+S_E(p)S_E(p-r)T(p,p^\prime;r), \label{phidef}
\end{equation}
where $2\gamma_-S_E(p)\equiv \gamma_-S(p)\gamma_-$, and $T_{\alpha\beta;\gamma\delta}(p,p^\prime;r)=(\gamma_-)_{\alpha\gamma}(\gamma_-)_{\beta\delta}T(p,p^\prime;r)$. The relation~\eqref{phidef} can be inverted, and the full vertex obtained in terms of $\phi(p,p^\prime;r)$ via
\begin{equation}
  T(p,p^\prime;r) = \frac{ig^2}{(p_- - p_-^\prime)^2} + \frac{i g^2 N_c}{\pi^2} \int \mathrm{d}k_- \frac{\phi(k_-,p^\prime;r)}{(k_- - p_-)^2}.  \label{Tphi}
\end{equation}

The function $\phi(p_-,p^\prime_-;r)$ satisfies the $\lambda$-independent Bethe-Salpeter equation (for the full derivation see~\cite{callan1976twodimensional,einhorn1976confinement}),
\begin{equation}
\mu^2\phi(x,x^\prime;r)=\frac{\pi^2}{N_c r_-(x-x^\prime)^2}+\alpha\left(\frac{1}{x}+\frac{1}{1-x}\right)\phi(x,x^\prime;r)+\int_0^1\frac{\left[\phi(x,x^\prime;r)-\phi(y,x^\prime;r)\right]\mathrm{d}y}{(x-y)^2},\nonumber\\
\end{equation}
where $\mu^2 = r^2\pi/g^2 N_c$, $\alpha=m^2\pi/g^2N_c$, and we switched to the more convenient variables $x=p_-/r_-$ and $x^\prime=p^\prime_-/r_-$.  Intuitively, here $x$ can be thought of as the share of the total light cone momentum carried by the quark within the meson. The full solutions $\phi(x,x';r)$ can be expressed in terms of solutions of the homogeneous equation
\begin{equation}
\mu^2\phi(x)=\alpha\left(\frac{1}{x}+\frac{1}{1-x}\right)\phi(x)+\int_0^1\frac{[\phi(x)-\phi(y)]\mathrm{d}y}{(x-y)^2}.\label{phix}
\end{equation}
(Here we suppress the $x^\prime$ and $r$ arguments.) Equation~\eqref{phix} has a discrete set of solutions~\cite{thooft1974twodimensional}, labeled by the integer $k$ and denoted $\mu_k$, $\phi_k(x)$. These solutions were examined numerically by 't~Hooft in~\cite{thooft1974twodimensional}; here we need only knowledge of some of their interesting properties. The label $k$ corresponds to the discrete spectrum of meson particles, bound states of a quark-antiquark pair. The solutions $\phi_k(x)$ form an orthonormal set, such that,
\begin{equation}
\sum_k\phi_k(x)\phi_k(x^\prime)=\delta(x-x^\prime), \qquad \int_0^1\phi_n^*(x)\phi_m(x)\mathrm{d}x=\delta_{nm}.
\end{equation}
For large values of $k$, so-called heavy meson bound states, the solutions can be approximated as
\begin{equation}
\mu_k^2\approx \pi^2 k,\qquad \phi_k(x)\approx\sqrt{2}\sin(\pi kx).
\end{equation}
The full wave function can be expressed in terms of the homogeneous solutions, $\phi_k(x)$, as
\begin{equation}
\phi(x,x^\prime;r)=-\sum_k\frac{\pi g^2}{(r^2-r_k^2)r_-}\,\int_0^1\mathrm{d}y\frac{\phi_k(x)\phi_k^*(y)}{(y-x^\prime)^2}.
\end{equation}
Combined with eq.~\eqref{Tphi}, it is then possible to recover the full quark-quark scattering amplitude, $T_{\alpha\beta,\gamma\beta}(p,p^\prime,r)$, see \cite{thooft1974twodimensional}. It is worth mentioning that the scattering amplitude contains $1/\lambda$ terms, which would diverge in the $\lambda\to0$ limit. This accounts for the fact that it is impossible for a meson state to decay into free quarks, and also, as we will later see, this ensures that the scattering amplitudes between mesons are finite.

To summarize the main results quoted in this section, in the large-$N_c$ limit of QCD in (1+1)-d, one can compute exactly the quark propagator and the quark-quark interaction vertex. The quark propagator exhibits a pole at $p\to\infty$, which is explained by the fact that the quark is confined. From the quark-quark interaction vertex one sees that only a discrete set of quark-antiquark configurations are allowed solutions, labeled by an integer $k$. These allowed quark-antiquark wave functions are associated with the discrete spectrum of mesonic bound states.

\subsection{Meson interactions}
\label{sec:mesonints}

The spectrum of the 't~Hooft model consists of an infinite tower of meson bounds states, with discrete mass values. Quarks and gluons are confined and so absent from the physical spectrum. With knowledge of the quark propagator and of quark-quark interactions, it is possible to compute the leading contributions to mesonic scattering amplitudes at large $N_c$.
\begin{figure}
  \begin{center}
    \includegraphics[width=0.6\textwidth]{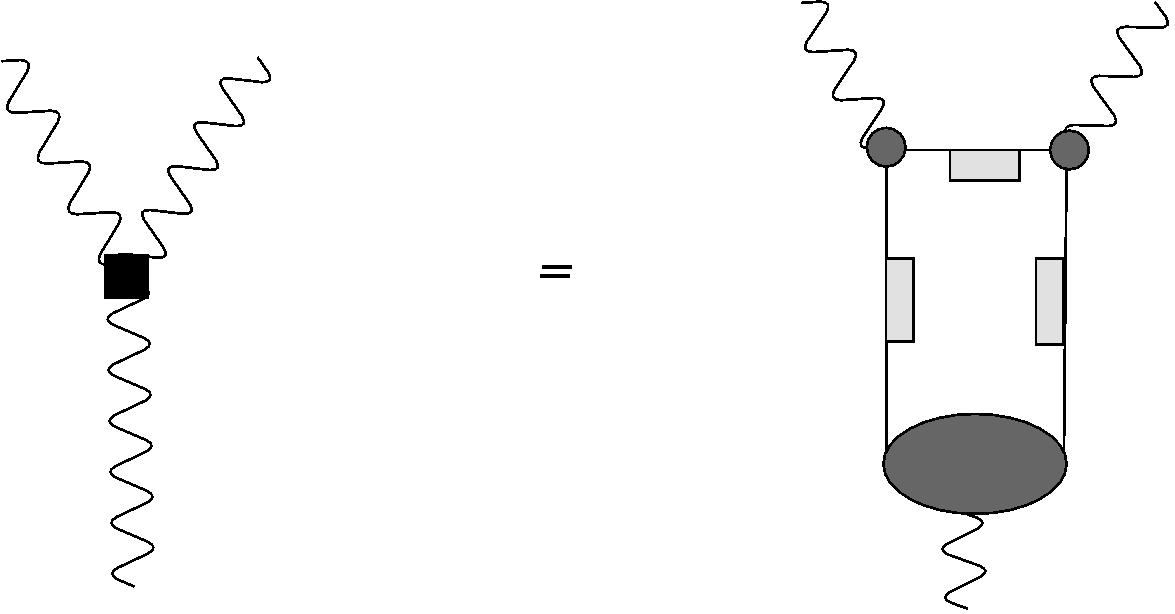}
  \end{center}
  \caption{The leading contribution to the $1\to2$ meson decay process. Mesons are represented by a wiggly line, and the new effective three-meson vertex is depicted as a black square blob. The ellipsoidal dark gray blobs represent quark-quark vertices, and the gray rectangular boxes represent full quark propagators.}
  \label{fig:mesonDecay}
\end{figure}
The simplest interaction, first discussed in \cite{callan1976twodimensional}, is the amplitude for $1\to2$ meson decay. As shown diagrammatically in figure~\ref{fig:mesonDecay}, this amplitude can be written in terms of the known quark propagator~\eqref{propagator} and the interaction vertex~\eqref{vertex}.

\begin{figure}
  \begin{center}
     \includegraphics[width=0.5\textwidth]{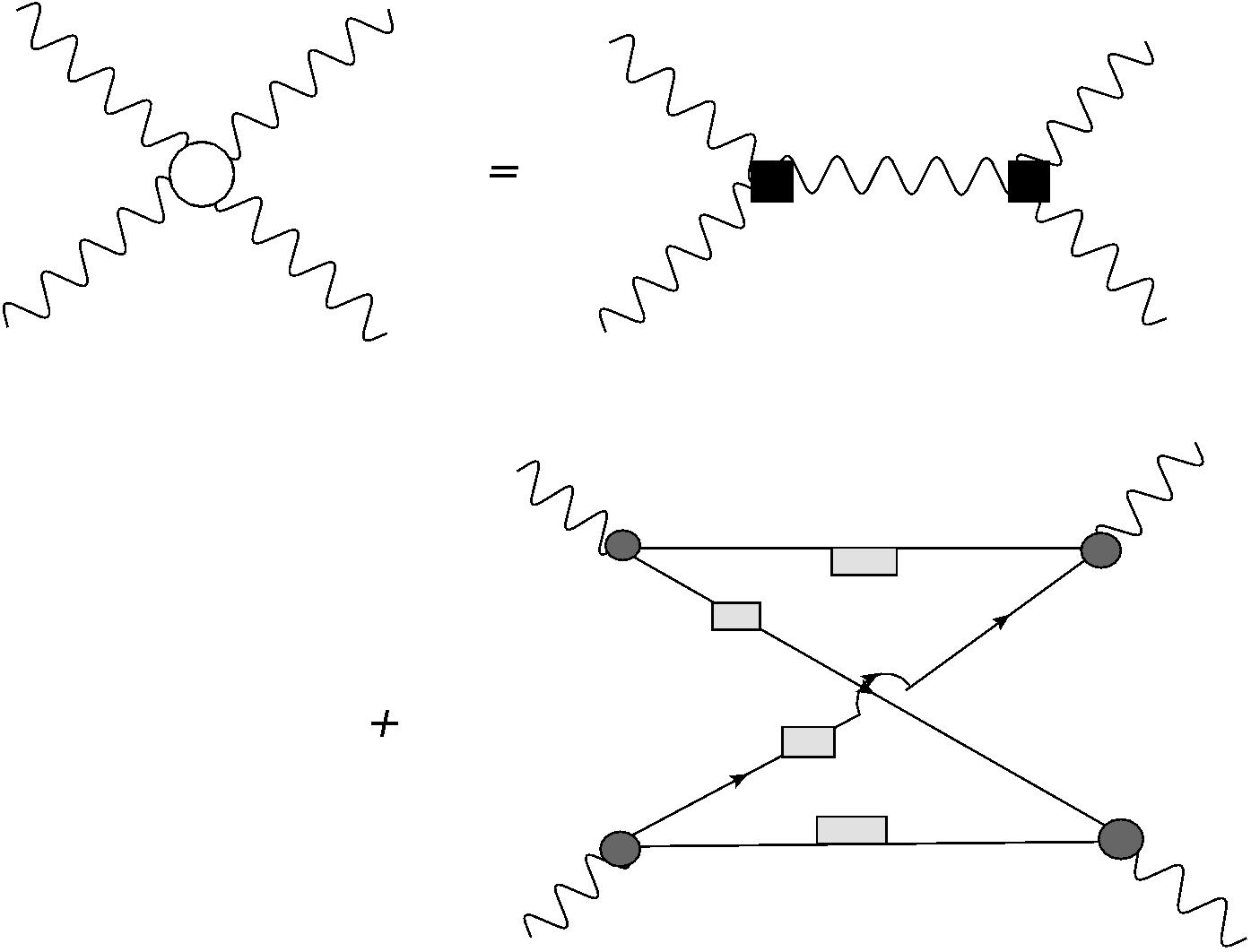}
  \end{center}
  \caption{A schematic depiction of the $2\to2$ meson scattering amplitude. The effective 4-meson vertex is represented by a white circular blob.}
  \label{fig:2to2}
\end{figure}

The factors of $\lambda$ that appear in the quark propagator, quark vertices, and arise from the integration over the loop momentum, exactly cancel to give a finite meson decay amplitude. We quote the final result for the amplitude (derived in \cite{callan1976twodimensional}) describing the decay of a single meson (described by meson quantum number $k_1$ and momentum $r_1$) into two mesons (with quantum numbers $k_2$ and $k_3$ and momenta $r_2$ and $r_3$),
\begin{align}
  A_{k_1;k_2,k_3}(r_1;r_2,r_3)=&
       \frac{4g^2\sqrt{N_c}}{\sqrt{\pi}}\left[\int_0^{r_{2-}}\mathrm{d}l_-\,\phi_1\left(\frac{l_-}{r_{1-}}\right)\phi_2\left(\frac{l_-}{r_{2-}}\right)\int_0^{r_{3-}}\mathrm{d}p_-\frac{\phi_3\left(\frac{p_-}{r_{3-}}\right)}{[p_--(l_--r_{2-})]^2}\right.\nonumber\\
 &\left.-\int_{r_{2-}}^{r_{1-}}\mathrm{d}l_-\,\phi_1\left(\frac{l_-}{r_{1-}}\right)\phi_3\left(\frac{l_--r_{2-}}{r_{3-}}\right)\int_0^{r_{2-}}\mathrm{d}p_-\frac{\phi_2\left(\frac{p_-}{r_{2-}}\right)}{(p_--l_-)^2}\right].
\end{align}
Here it is important to notice that this amplitude is $\mathcal{O}(1/\sqrt{N_c})$. Thus, in the strict $N_c\to\infty$ limit, the mesons are completely stable and do not decay. Similarly, one can study the $2\to2$ meson scattering amplitude in terms of the exact quark propagator and interaction vertex. This is shown diagrammatically in figure~\ref{fig:2to2}. Its leading contribution is at order $\mathcal{O}(1/N_c)$. 

As we have seen, non-trivial meson scattering involving three particles happens at order $1/\sqrt{N_c}$, while scattering involving four particles happens at order $1/N_c$. It is easy to see a pattern: when higher numbers of particles are involved, the leading contribution to non-trivial scattering comes at higher orders of $1/N_c$. A non-trivial interaction vertex involving $n$ particles will only contribute at order $1/N_c^{(n-2)/2}$.  As we will explore in the next section, there is a deep correspondence between powers of $1/N_c$ and the number of particles involved in a process. This can also be seen at the level of matrix elements of local operators. 

\section{\boldmath Particle number/power of $N_c$ correspondence}
\label{sec:correspondence}

We have seen that one can diagrammatically compute the leading contributions to scattering amplitudes under a $1/N_c$ expansion. This approach can also be used to study matrix elements of local operators, which are necessary ingredients when studying the non-equilibrium dynamics of such systems. Physical observables have to be gauge invariant operators,  for instance we can construct a meson-like operator that is given by an quark and antiquark pair connected by a path-ordered Wilson line,
\begin{equation}
S_\gamma(x,y)=\bar{q}_j(x) W^{ji}_{\gamma}(x,y)q_i(y),
\end{equation}
where $W^{ji}_{\gamma}(x,y)$ is a path-ordered exponential of the gauge field along a path, $\gamma$, which has endpoints at $x$ and $y$,
\begin{equation}
W^{ji}_{\gamma}(x,y)=\mathcal{P}\exp\bigg[\int_\gamma \mathrm{d}x^\mu A^{ji}_\mu(x)\bigg].
\end{equation}
We call operators such as $S(x,y)$ ``mesonic operators", because they are gauge invariant, bilinear in the quark field $q_i$, and create meson states when applied to the ground state.

We can also define gauge-invariant ``gluonic operators", which do not contain the quark field. These are defined in terms of the Wilson loop, given by a Wilson line along a closed loop $\mathcal{C}$
\begin{equation}
W[\mathcal{C}]={\rm Tr}\,\mathcal{P} \exp\bigg[\oint_\mathcal{C} dx^\mu A^{ji}_\mu(x)\bigg].
\end{equation}

For the sake of simplicity, in our discussion we will focus on local mesonic operators, where the quark and antiquark are at the same point,
\begin{equation}
  S(x)=\bar{q}_i(x)q^i(x). \label{observable}
\end{equation}
When computing correlation functions of this operator, we will need matrix elements of the form $\langle 0\vert S(x)\vert n_1,p_1;n_2,p_2;\dots\rangle$, which are evaluated between the ground state $|0\rangle$, and a state with a set of mesons labeled by quantum numbers $n_i$ and momenta $p_i$.

\begin{figure}
  \begin{center}
    \includegraphics[width=0.4\textwidth]{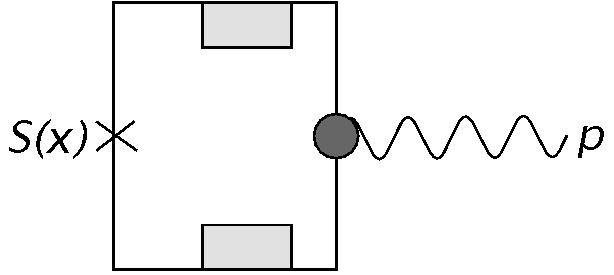}
  \end{center}
  \caption{Diagrammatic depiction of the one-meson matrix element.}
  \label{fig:oneMeson}
\end{figure}

The zero-meson matrix element (vacuum expectation value) has been computed in \cite{zhitnitsky1985chiral} (see also~\cite{li1986largen,lenz1991hamiltonian}) and is given (in the $m\to0$ limit) by~\footnote{Note that in the light cone gauge this result is not trivial to obtain, being subtly related to boundary conditions and zero modes. Refs.~\cite{zhitnitsky1985chiral} and~\cite{li1986largen} work in a different gauge to avoid this issue. It is also subtle in that it signals a condensate of the mesons, which breaks the (continuous) chiral symmetry. This should be forbidden by the Mermin-Wagner-Hohenberg/Coleman theorem~\cite{mermin1966absence,hohenberg1967existence,coleman1973there}, but the $N_c = \infty$ limit removes the self-interactions of the Goldstone modes that would usually leads to an infrared divergence and force $\langle 0 | S(x) | 0 \rangle = 0$.} 
\begin{equation}
\langle 0\vert S(x)\vert 0\rangle=-\frac{N_c}{\sqrt{12}}\left(\frac{N_cg^2}{2\pi}\right)^{1/2} = \mathcal{O}\left(N_c\right).\label{vev}
\end{equation}
For finite bare quark mass $\langle 0 | S(x) | 0\rangle$ remains ${\cal O}(N_c)$, see ref.~\cite{burkardt1994exact}. It was shown in \cite{callan1976twodimensional} that the matrix elements with mesons can be computed diagrammatically.  The diagram corresponding to the one-meson matrix element is given in figure~\ref{fig:oneMeson}. This yields the one-particle matrix element
\begin{equation}
  \langle0\vert S(x)\vert n,p\rangle =
  \left(\frac{N_c}{\pi}\right)^{1/2}
  \int_0^1\mathrm{d}y\, m\left(\frac{1}{y}-\frac{1}{1-y}\right)\phi_n(y)
  =\mathcal{O}(\sqrt{N_c}).
\end{equation}

For the purposes of this paper, the exact details of the matrix elements with higher number of mesons are not important: we only need to point out the fact that there is a correspondence between the number of mesons in the matrix elements and the order of $1/N_c$ of its leading contribution. In fact, it can be shown diagrammatically that
\begin{align}
\langle 0\vert S(x)\vert n_1,p_1;n_2,p_2;\dots;n_k,p_k\rangle=\mathcal{O}\left(\frac{1}{N_c^{(k-2)/2}}\right).\label{particlencorrespondence}
\end{align}
Matrix elements with mesons in both the incoming and outgoing state can be obtained from \eqref{particlencorrespondence} by using crossing symmetry. Thus the correspondence \eqref{particlencorrespondence} is still true, connecting the total number of particles in incoming and outgoing states to the power of $1/N_c$.

One can take advantage of this correspondence, as in~\cite{callan1976twodimensional}, to compute correlation functions up to some given power of $1/N_c$. For example, one can compute the leading contributions to 
\begin{align}
M(q^2)=\int \mathrm{d}x \,e^{{\rm i}q\cdot x}\langle 0\vert \mathcal{T}\big(S(x)S(0)\big)\vert 0\rangle
\end{align}
by inserting a complete set of meson states between the two operators (obtaining the Lehmann representation). To leading order in $1/N_c$ only the one-meson contribution is relevant, such that
\begin{align}
M(q^2)=\sum_n\frac{\left\vert\langle 0\vert S(0)\vert n,0\rangle\right\vert^2}{q^2-\mu_n^2}+\mathcal{O}\left(\frac{1}{N_c^2}\right),\label{onemesonexpansion}
\end{align}
from which the expected behavior characteristic of asymptotic freedom can be recovered,
\begin{align}
  M(q^2)\sim \left(\frac{N_c}{\pi}\right) \ln\left(\frac{q^2}{\mu^2}\right)\quad \text{as} \quad q\to\infty,
\end{align}
as discussed in detail in Ref.~\cite{callan1976twodimensional}.

It is important to point out that the expansion~\eqref{onemesonexpansion} is not a low-energy approximation for the correlation function, even though it only includes contributions from single mesons. As a result, one can study the ultraviolet properties of correlation functions, such as asymptotic freedom, to arbitrarily high energy scales provided the power of $1/N_c$ is fixed. 

In the  Section \ref{sec:quenches} we will see how this deep connection between the number of mesons and powers of $1/N_c$ has strong consequences for the thermalization properties of the 't~Hooft model.

\section{Large $N_c$ volume independence}
\label{sec:volumeindependence}

One of the most powerful, and surprising, results in large-$N_c$ gauge theories is the phenomenon of volume independence~\cite{eguchi1982reduction,kovtun2007volume}. Consider a gauge invariant operator $O$ of a confining pure gauge theory (with no fermions) in $d$ space-time dimensions, with one of the spatial dimensions compactified into a circle of circumference $\beta$. The statement of large-$N_c$ volume independence is 
\begin{equation}
\langle O\rangle_\beta=\langle O\rangle_{\infty}\left[1+\mathcal{O}\left(\frac{1}{N_c^2}\right)\right],
\end{equation}
where $\langle \dots\rangle_\beta$ represents the ground-state average of an operator in the space-time geometry $R^{d-1}\times S^1$.  This statement is true as long as $\beta$ is larger than any critical value $\beta_c$, where the system may enter a deconfined phase. That is, as long as the system remains in a confined phase, expectation values of gauge invariant observables remain insensitive to the compactification of one of the spatial dimensions, up to $\mathcal{O}(1/N_c)$ corrections.

\subsection{Thermal equilibrium}

The existence of  volume independence in a Lorentz invariant theory has immediate consequences for thermodynamics. If instead of a spatial dimension, we compactify a Euclidean time dimension to a circumference $\beta$, then that is equivalent to gauge-invariant observables being temperature independent:
\begin{equation}
\langle O\rangle_\beta=\frac{{\rm Tr}\,\left[O\,e^{-\beta H}\right]}{{\rm Tr} \,e^{-\beta H}}=\langle0\vert O\vert 0\rangle\left[1+\mathcal{O}\left(\frac{1}{N_c^2}\right)\right],\label{thermalindependence}
\end{equation}
as long as one does not reach a deconfining critical temperature $\beta_c$. We will see in the following that this statement can be modified to gauge theories including fermions, but the power of the finite-temperature corrections are modified.

The implications of thermal independence at large $N_c$ can be understood by considering the factorization properties of correlation functions of gauge-invariant operators.  Let  $O_1,\,O_2$ be two such operators, then their ground-state correlation functions satisfy 
\begin{equation}
\langle O_1O_2\rangle=\langle O_1\rangle\langle O_2\rangle\left[1+\mathcal{O}\left(\frac{1}{N_c^\alpha}\right)\right],\label{factorizationst}
\end{equation}
where the connected piece of the correlation function is suppressed at large $N_c$. The power, $\alpha$, depends on whether the operators $O_1$, $O_2$ are mesonic or gluonic.

We can, in fact, write down the general $1/N_c$ power of the leading contribution to multi-point connected correlation functions. Given a gauge theory with both quarks and gluons, we define $O^g_i$ to be purely gluonic operators (such as the Wilson loop $W[\mathcal{C}]$) and $O^f_i$ to be mesonic operators (such as $S(x)$). Furthermore, we impose that these gauge invariant operators are non-factorizable, i.e. they cannot be decomposed into a product of other gauge invariant operators (for instance, $S^2(x)$ is not allowed). We normalize these operators such that
\begin{align}
\langle O^f_1 O^f_2\rangle\sim N_c^0,\qquad \langle O^g_1O^g_2\rangle\sim\frac{1}{N_c}.
\end{align}
It can then be shown that the leading contributions to connected correlation functions under a large $N_c$ expansion are~\cite{Manohar:1998xv},
\begin{align}
\langle O^g_1\dots O^g_nO^f_1\dots O^f_m\rangle_{\text{connected}}\sim \mathcal{O}\left(\frac{1}{N_c^{n+m/2-1}}\right).
\end{align}

In the thermodynamic limit, the thermal expectation value~\eqref{thermalindependence} is equivalent to the microcanonical ensemble average, where the average is taken over states within a small energy window $\Delta E$ of the thermal energy $E_\beta$; recall eq.~\eqref{mce},
\begin{align}
  \langle O \rangle_{\text{MCE}} = \frac{1}{N_{\Delta E}}\sum_{\substack{i:\\|E_i - E_\beta|<\Delta E}} \langle E_i | O | E_i \rangle.\label{mceagain}
\end{align}
If $\beta$ is such that the system is still in a confined phase, then all the states $\vert E_i\rangle$ are color-singlets. This implies that we can create all such states by acting on the ground state with a non-factorizable gauge invariant operator (or a product of them), $\hat{E}_i$, such that
\begin{align}
\vert E_i\rangle=\hat{E}_i\vert 0\rangle, \qquad \langle \hat{E}_i\rangle=0.
\end{align}
We can then express the microcanonical ensemble result for the correlation function as a sum of correlation functions of the form 
\begin{align}
\langle \hat{E}_i O\hat{E}_i\rangle=\langle O\rangle \langle \hat{E}_i\hat{E}_i\rangle+\langle \hat{E}_i O\hat{E}_i\rangle_{\text{connected}},\label{eoe}
\end{align}
where we have used the factorization (\ref{factorizationst}). The sum over index $i$ in the first, disconnected term of eq.~\eqref{eoe} yields the vacuum expectation value of $O$, which is of order $N_c^0$. 

The connected term is suppressed as
\begin{align}
\langle \hat{E}_i O\hat{E}_i\rangle_{\rm connected}\sim \frac{\langle O\rangle}{N_c^{\alpha_i}},\nonumber
\end{align}
where the power, $\alpha_i$, depends on whether the theory contains fermionic fields or is a pure gauge theory. In a purely gluonic gauge theory both $\hat{E}_i$ and $O$ are necessarily gluonic and are, at most, non-factorizable operators such that $\alpha_i=2$. This recovers the known result for large-$N_c$ volume independence~\cite{eguchi1982reduction}, eq. (\ref{thermalindependence}). In the case where the theory contains fermions, the leading contributions will come from $\hat{E}_i$ being a mesonic, non-factorizable operator and as a result $\alpha_i=1$.

As long as $\beta$ stays within a confining phase~\footnote{The separation between a deconfined and a confined phase is not straightforward in gauge theories containing massive fermions. There are different order parameters, such as the quark condensate and Polyakov loops (see for instance~\cite{Hatta:2003ga}), whose expectation values differentiates between the two phases. The critical temperature, however, is different for the each order parameter, suggesting that there is a crossover, rather than a sharp transition, between the confined and deconfined phase. For our purpose, by confined phase we mean that all the states within the energy window $\vert E_i-E_\beta\vert<\Delta E$, are color singlets that can be created by products of non-factorizable operators.}, all thermal corrections to expectation values of gauge-invariant operators are suppressed by powers of $1/N_c$. We now want to understand if this phenomenon of volume-independence can be generalized from thermal equilibrium to non-equilibrium dynamics.

\subsection{Generalization to non-equilibrium dynamics}

A generic non-equilibrium setup involves preparing the system in an initial density matrix, $\rho_0$ that does not commute with the Hamiltonian of the system, $H$. Time-evolution is then described through the time-dependent density matrix 
\begin{align}
\rho(t)=e^{-{\rm i}Ht}\rho_0\,e^{{\rm i}Ht},
\end{align}
and the evolution of physical observables is computed via
\begin{align}
\langle O(t)\rangle=\frac{{\rm Tr}\left[O \,\rho(t)\right]}{{\rm Tr}\rho_0}.
\end{align}
Here the trace can taken by summing over eigenstates of the Hamiltonian, $\vert E_i\rangle$. The non-trivial time-dependence arises from the fact that the density matrix is, generally, not diagonal in this eigenstate basis. Physical observables can then be expressed as
\begin{align}
\langle O(t)\rangle=\frac{\sum_{i,j} \rho_{ij} \,e^{-{\rm i}t(E_i-E_j)} \langle E_j\vert O\vert E_i\rangle}{\sum_i \rho_{ii}},
\end{align}
where $\rho_{ij}=\langle E_i\vert \rho_0\vert E_j\rangle$. 

We now want to examine the analogue of volume independence in these non-equilibrium scenarios, where the system remains in the confined phase. This can be defined precisely by assuming that the initial conditions are such that $\rho_{ij}$ is non-zero only for color-singlet states, $\vert E_i\rangle,\,\vert E_j\rangle$. We assume the matrix $\rho_{i,j}$ vanishes when evaluated on colorful states above the deconfinement threshold. 

Given that the sum $\sum_{i,j}$ runs only over color-singlet states (which can be created by products of color-singlet, non-factorizable operators) we arrange the contributions to the time-dependent expectation value by their $1/N_c$ dependence
\begin{eqnarray}
\langle O(t)\rangle\sum_i\rho_{ii}&=&\langle O\rangle+\sum_{i\neq0}\Big(\rho_{0i} e^{{\rm i}tE_i}\langle E_i\vert O\vert 0\rangle+\rho_{i0}e^{-{\rm i}tE_i}\langle 0\vert O\vert E_i\rangle\Big)\nonumber\\
&&+\sum_{i,j\neq0}\rho_{ij} e^{-{\rm i}t(E_i-E_j)}\langle E_j\vert O\vert E_i\rangle.
\end{eqnarray}
The leading contribution comes from the diagonal (time-independent) term $\langle O\rangle$. The rest of the terms are suppressed as
\begin{align}
 \langle E_i\vert O\vert 0\rangle\sim\langle 0\vert O\vert E_i\rangle\sim\frac{\langle O\rangle}{N_c^{\alpha_1}},\qquad\langle E_j\vert O\vert E_i\rangle\sim\frac{\langle O\rangle}{N_c^{\alpha_2}}.\label{twopowers}
 \end{align} 

If we consider a purely gluonic theory, the leading contributions come from when $\hat{E}_i,\,\hat{E}_j,\, O$ are non-factorizable gluonic operators such that $\alpha_1=1$ and $\alpha_2=2$. If the theory contains fermions, then the leading contributions come from mesonic, non-factorizable operators $\hat{E}_i,\,\hat{E}_j$, such that $\alpha_1=1/2$ and $\alpha_2=1$. It is then clear that non-equilibrium dynamics affects observables only at a sub-leading orders in the $1/N_c$ expansion, generalizing the concept of volume independence. In thermal-equilibrium, the corrections come at order $\langle O\rangle/N_c^2$ in pure gauge theories, or at $\langle O\rangle/N_c$ for theories with quarks. In the general non-equilibrium case, however, the highly non-diagonal matrix elements, $\langle 0\vert O\vert E_i\rangle$ bring new lower-order corrections, at order $\langle O\rangle/N_c$ for pure gauge theories and at  $\langle O\rangle/\sqrt{N_c}$ for gauge theories with quark.

We point out that  similar observations regarding the non-equilibrium manifestations of large-$N_c$ volume independence had already been made in Ref. \cite{Cubero:2016enh}, for the case of a quantum quench of the principal chiral sigma model (PCSM). The PCSM is not a confining gauge theory, it is in fact an integrable field theory with an $SU(N_c)_L\times SU(N_c)_R$ global chiral symmetry, and with (matrix-valued) particle excitations, correspondingly labeled with left and right $SU(N_c)$ color indices. The arguments presented in this section are therefore not expected to apply to the PCSM. However, when one performs quantum quenches from initial states which are $SU(N_c)_L\times SU(N_c)_R$ color singlets, it was shown that contributions to the non-equilibrium dynamics are, analogously to eq.~\eqref{twopowers}, suppressed by two different powers of $1/N_c$. This feature, combined with the integrability of the PCSM was used in \cite{Cubero:2016enh} to compute the full time evolution of local observables, up to the leading order in $1/N_c$.

\section{Quenches from multi-meson states}
\label{sec:quenches}

In recent works~\cite{james2019nonthermal,robinson2019signatures} (see also~\cite{kormos2016realtime,rakovszky2016hamiltonian,hodsagi2018quench}) it was shown that for the quantum Ising chain with both transverse and longitudinal fields, the existence of a tower of meson states, stretching far into the many-body spectrum, leads to a violation of the eigenstate thermalization hypothesis. This is due to the fact that one can create eigenstates with a finite number (e.g., one) of mesons that carry an extensive amount of energy. Then, when a physical observable is evaluated in such a states, its expectation value does not agree with the (microcanonical) thermal expectation value at the same energy density. This violation of the eigenstate thermalization hypothesis has consequences for non-equilibrium dynamics: if one performs a quantum quench in which the initial state projects strongly onto these extensive-energy few meson states, the system does not thermalize following time-evolution. 

In this section we will explore a similar scenario in the 't~Hooft model, by considering a quantum quench from an initial state with a finite number of mesons (i.e., zero density), but with extensive energy. The expectation values following this quench will be compared to the expected value at the corresponding finite temperature.

\subsection{Thermal expectation value}
\label{sec:thermal}
Given the results of the previous sections, we can straightforwardly compute thermal expectation values of local operators, at leading order in $1/N_c$. For the di-quark operator,~\eqref{observable}, the thermal average is given as usual by
\begin{equation}
\langle S(x)\rangle_\beta={\rm Tr}\left[e^{-\beta H} S(x)\right]/Z,\qquad Z={\rm Tr}\left[ e^{-\beta H}\right].
\end{equation}
The trace can be computed in the basis of meson states. In this basis, the Hamiltonian can be written effectively as \cite{thooft1974planar} (see also~\cite{lenz1991hamiltonian})
\begin{equation}
H=H_0 +\frac{1}{\sqrt{N_c}} H_{\rm int},\qquad H_0=\int \mathrm{d}p \sum_nH^0_n(q) \,A_n^\dag(p)A_n(p)\label{hamiltonian}
\end{equation}
where $A^\dag_n(p)$ is the creation operator for a meson with quantum number $n$, and momentum $p$. The meson-meson interactions come at higher orders in $1/\sqrt{N_c}$, as we saw in section~\ref{sec:mesonints}.

To leading order in $1/N_c$, the thermal expectation value is computed by considering the contributions from one-meson states only
\begin{equation}
\langle S(x)\rangle_\beta=\langle0\vert S(x)\vert 0\rangle+\sum_n\int \mathrm{d}p \,W(n,p) \langle n,p\vert S(x)\vert n,p\rangle_{\text{connected}}+\mathcal{O}\left(\frac{1}{N_c}\right),\label{thermalexpectation}
\end{equation}
with the first term on the right-hand side being the vacuum expectation value, which is of order $N_c$, as seen in eq.~\eqref{vev}. The matrix elements involving two mesons are of order 1 (suppressed, as we expect from large-$N_c$ volume independence) and $W(n,p)$ is the Bose-Einstein thermal distribution for each type of meson. Here the connected matrix element, $\langle n,p\vert S(x)\vert n,p\rangle_{\text{connected}}$, denotes that we have subtracted disconnected contributions, such as $\langle 0\vert S(x)\vert 0\rangle \langle n,p\vert n,p\rangle.$ 

With an explicit expression for the thermal expectation value of $S(x)$ at hand,~\eqref{thermalexpectation}, we can test if time-evolution of a given initial state with the same energy density will reproduce the thermal result, to order $N_c^0$, at late times. 

\subsection{Expectation value after multi-meson quench}
Accordingly, we consider an initial state given by a finite set of stationary heavy mesons,
\begin{equation}
\vert \psi_0\rangle=\vert n_1,0;n_2,0;\dots;n_k,0\rangle,\label{initialstate}
\end{equation}
chosen such that the total energy $E=\sum_{i=1}^k \mu_{n_i} \sim L$ is extensive and agrees with that of the thermal state for a given temperature, $\beta$. The state~\eqref{initialstate} is not an eigenstate of the full Hamiltonian, eq.~\eqref{hamiltonian}, being instead an eigenstate of only the non-interacting (first) term. The state will therefore evolve non-trivially under $\vert\psi_t\rangle=e^{-{\rm i}tH}\vert \psi_0\rangle$. We are interested in computing the time-dependent expectation values of local observables, such as
\begin{equation}
\langle S(x)\rangle_t=\frac{\langle \psi_0\vert e^{{\rm i}tH}S(x)e^{-{\rm i}tH}\vert \psi_0\rangle}{\langle \psi_0\vert\psi_0\rangle}.
\end{equation}
It is convenient to express this in the interaction picture,
\begin{equation}
\langle S(x)\rangle_t=\frac{\langle \psi_0\vert U_I^\dag(t)S_I(x,t)U_I(t)\vert \psi_0\rangle}{\langle \psi_0\vert\psi_0\rangle},\label{interactionpicture}
\end{equation}
where $S_I(x,t)=e^{{\rm i} tH_0}S(x)e^{-{\rm i}tH_0}$, $U_I(t)=e^{{\rm i}tH_0}e^{-{\rm i}tH}$ and $H_0$ is given in eq.~\eqref{hamiltonian}. It is then possible to expand the interaction picture time-evolution operator $U_I(t)$ as a perturbative series in powers of $1/N_c$.

The leading contribution to the perturbative expansion of \eqref{interactionpicture} is obtained by taking $U_I(t)\approx 1$. This gives the time-independent expectation value
\begin{equation}
\langle S(x)\rangle_t=\langle 0\vert S(x)\vert 0\rangle+\sum_{i=1}^k\frac{\langle n_i,0\vert S(x)\vert n_i,0\rangle_c}{\langle n_i,0\vert n_i,0\rangle}+\mathcal{O}\left(\frac{1}{N_c}\right).\label{perturbative}
\end{equation}
Here the first term (involving the matrix element with zero-meson states) is of order $N_c$ and the second term (with one-meson states) is order $1$.

Comparing eq.~\eqref{perturbative} with eq.~\eqref{thermalexpectation}, it is immediately clear that up to order $N_c^0$ the two do not necessarily agree. One could be tempted, therefore, to conclude that the eigenstate thermalization hypothesis is not satisfied for the initial state~\eqref{initialstate}. It is, however, too early to reach this conclusion.

As we have previously discussed, the onset of thermalization is a non-perturbative phenomenon. Integrable systems reach non-thermal steady states at late times, which are at a distance $\mathcal{O}(1)$ from thermal values~\footnote{See, e.g., ref.~\cite{essler2014quench} for a discussion of this in the context of time-evolution in the presence of weak integrability-breaking perturbations.}. At first glance it would seem that the perturbative expansion of $\langle S(x)\rangle_t$, based on the expansion of the time evolution operator $U_{I}(t)$, in powers of $1/N_c$, can only generate small subleading corrections to eq.~\eqref{perturbative}.

We will see, however, that it is still possible (in principle) to obtain additional $\mathcal{O}(1)$ corrections  from the perturbative expansion of $U_I(t)$. The only way this can happen is if some other quantity in the expansion diverges as $1/N_c$ is taken to 0, such that the terms in the expansion are not actually ``small'', but instead $\mathcal{O}(1)$. We will see that this is indeed the case, as thermalization is expected to occur only at sufficiently late times, such that $t^\alpha\sim N_c\to\infty$, for some positive power $\alpha$.

\subsubsection{Generating non-perturbative corrections in the long-time limit}

With the aim of finding $\mathcal{O}(1)$ corrections to eq.~\eqref{perturbative}, we are  interested in computing the leading contributions as $t\to\infty$ from the $1/N_c$ corrections. In order for this expansion to produce $\mathcal{O}(1)$ terms, we would need the $\mathcal{O}(1/N_c)$ correction to include a term that grows with a positive power of time, such that at late times, the expectation value would be of the form
\begin{equation}
\langle S(x)\rangle_t=\langle 0\vert S(x)\vert 0\rangle+\sum_{i=1}^k\frac{\langle n_i,0\vert S(x)\vert n_i,0\rangle_c}{\langle n_i,0\vert n_i,0\rangle}+\frac{t^{\alpha_1}}{N_c}A_1+\frac{t^{\alpha_2}}{N_c^2}A_2+\dots,\label{growingterms}
\end{equation}
where $\alpha_1,\alpha_2,\dots$ are positive numbers. A simple example of how such terms that grow with time can produce $\mathcal{O}(1)$ contributions is, for instance, if it happens that $\alpha_n=n$ and $A_n=(-1)^n\frac{Ab^n}{n!}$, in which case the expansion exponentiates at late times:
\begin{equation}
\langle S(x)\rangle_t=\langle 0\vert S(x)\vert 0\rangle+\sum_{i=1}^k\frac{\langle n_i,0\vert S(x)\vert n_i,0\rangle_c}{\langle n_i,0\vert n_i,0\rangle} -A+Ae^{-\frac{b}{N_c}t}.\label{expthermalization}
\end{equation}
In this scenario of the series exponentiating at late times, the perturbative expansion generates a non-perturbative term, $A$, which in principle could move $\lim_{t\to\infty}\langle S(x)\rangle_t$ towards the thermal value $\langle S(x)\rangle_\beta$. We point out that a very similar scenario of exponentiation leading to non-perturbative late-time behavior has been seen in the case of a quantum quench of the Ising field theory~\cite{schuricht2012dynamics} (and the corresponding lattice model~\cite{calabrese2011quantum,calabrese2012quantum,calabrese2012quantum2}) and the sine-Gordon model~\cite{bertini2014quantum,cubero2017quantum,horvath2018overlap}.

\subsubsection{Absence of $t\to\infty$ non-perturbative corrections at $\mathcal{O}(1/N_c)$}

We will now show that it is not possible to generate the non-perturbative corrections that would lead to thermalization in the 't~Hooft model. To do so, we need to understand the structure of the matrix elements involved, and how it may be possible to produce terms that grow with time. 

Let us examine the time-dependence of the leading $\mathcal{O}(1/N_c)$ correction to eq.~\eqref{interactionpicture}; there are two new contributions at this order. The first comes from the $\mathcal{O}(1/N_c)$ contribution to $\langle \psi_0\vert S_I(x,t)\vert \psi_0\rangle$. This static term arises from matrix elements with states containing two particles, bringing a contribution of the form
\begin{equation}
\sum_{\substack{i,j=1;\\j\neq i}}^{k}\frac{\langle n_i,0;n_j,0\vert S(x)\vert n_i,0;n_j,0\rangle_c}{\langle n_i,0;n_j,0\vert n_i,0;n_j,0\rangle}.
\end{equation}
This is $\mathcal{O}(1/N_c)$, but time-independent and thus not of the form~\eqref{growingterms} , and will not contribute at late times.

The second, more relevant, contribution arises from the $\mathcal{O}(1/N_c)$ expansion of the time-evolution operator, $U_I(t)$. At this order in the expansion, we include the possibility for one of the mesons in the initial state to decay into a pair of mesons with opposite momenta (consistent with momentum conservation). This contribution to the expectation value is then
\begin{align}
\frac{1}{N_c}\sum_{i,j=1}^k\, \sum_{i^\prime,i^{\prime\prime}=1}^\infty \sum_{j^{\prime},j^{\prime\prime}=1}^\infty\int \mathrm{d}p_i \mathrm{d}p_j& \,\mathcal{A}^*_{j; j^{\prime},j^{\prime\prime}}(p_j) \,\mathcal{A}_{i;i^{\prime},i^{\prime\prime}}(p_i)\,e^{{\rm i}t(E_{i^{\prime},p_i}+E_{i^{\prime\prime},-p_i}-E_{j^{\prime},p_j}-E_{j^{\prime\prime},-p_j})} \nonumber\\
& \times\frac{\langle n_{j^{\prime}},p_j\vert S(x)\vert n_{i^{\prime}},p_i\rangle_c}{\langle n_j,0\vert n_i,0\rangle}\langle n_{j^{\prime\prime}},-p_j\vert n_{i^{\prime\prime}},-p_i\rangle. \label{contrib2}
\end{align}
Here $\frac{1}{\sqrt{N_c}}\mathcal{A}_{i;i^{\prime},i^{\prime\prime}}(p_i)$ is the amplitude for a meson with zero momentum and quantum number $n_i$ to decay into a pair of mesons with quantum numbers $n_{i^\prime}$ and $n_{i^{\prime\prime}}$ and momenta $p_i$ and $-p_i$ respectively. This decay process, as discussed above, is suppressed by a factor of $1/\sqrt{N_c}$, which  we have explicitly factored out for clarity.

The expression~\eqref{contrib2} is simplified by using the factor $\langle n_{j^{\prime\prime}},-p_j\vert n_{i^{\prime\prime}},-p_i\rangle$ to eliminate one of the momentum integrals and a summation over quantum numbers, yielding
\begin{align}
\frac{1}{N_c}\sum_{i,j=1}^k\,\sum_{i^{\prime},i^{\prime\prime},j^{\prime}=1}^\infty\int \mathrm{d}p_i&  \,\mathcal{A}^*_{j; j^{\prime},i^{\prime\prime}}(p_i) \,\mathcal{A}_{i;i^{\prime},i^{\prime\prime}}(p_i)\,e^{{\rm i}t(E_{i^{\prime},p_i}+E_{i^{\prime\prime},-p_i}-E_{j^{\prime},p_i}-E_{i^{\prime\prime},-p_i})} \nonumber\\
&\times\frac{\langle n_{j^{\prime}},p_i\vert S(x)\vert n_{i^{\prime}},p_i\rangle_c}{\langle n_j,0\vert n_i,0\rangle}. \label{timeEvo1overN}
\end{align}
The only time dependence present in~\eqref{timeEvo1overN} is in the oscillatory phase, \[\exp\left[{\rm i}t\left(E_{i^{\prime},p_i}+E_{i^{\prime\prime},-p_i}-E_{j^{\prime},p_i}-E_{i^{\prime\prime},-p_i}\right)\right].\] This appears within an integral over momentum $p_i$ and summations over quantum numbers. At late times, this generically becomes a rapidly oscillating phase and, since we are integrating over $p_i$ and summing over the quantum numbers $i^\prime,i^{\prime\prime},j^\prime$, by stationary phase arguments we find this contribution generally decays at late times as some power law in $t$.

The only way to obtain a behavior other than power-law decay is if there is some singularity in the integrand, specifically in the matrix elements $\langle n_{j_1},p_i\vert S(x)\vert n_{i_1},p_i\rangle$, which could give additional non-decaying, non-trivial contributions besides the stationary phase expectation. This has indeed been observed in similar calculations of the late-time dynamics in the Ising field theory and the sine-Gordon model, see~\cite{schuricht2012dynamics,calabrese2011quantum,calabrese2012quantum,calabrese2012quantum2,bertini2014quantum,cubero2017quantum,horvath2018overlap}. There singularities in the matrix elements lead to non-perturbative late-time dynamics.

At this point we note that there are \textit{significant differences} between matrix elements of the Ising field theory/the sine-Gordon model (which contain singularities that can produce growing terms at large times) and those that we consider here. In particular the matrix element $\langle n_{j_1},p_i\vert S(x)\vert n_{i_1},p_i\rangle_c$ does not contain any singularities that would produce any non-trivial terms. Indeed, it is well understood in the literature of integrable models in (1+1)-d where these singularities in the matrix elements come from. These are known as ``annihilation poles'' which occur when a particle in the outgoing state can annihilate with an identical particle in the incoming state, which leads generally to there being a simple pole when the momentum of the particles approach each other~\cite{smirnov1992form,mussardo:2010mgq}.

It is well known that annihilation poles are always present in matrix elements which contain two-or-more particles in both the incoming and outgoing states. On the other hand, matrix elements with only a single incoming and outgoing particle do not generally have such annihilation poles. Indeed, such matrix elements only contain annihilation poles in the case where the particles are \textit{non-local} with respect to the operator in question~\cite{delfino1996nonintegrable,mussardo2010statistical}, $S(x)$ in the case at hand. That is, when particles are topological kinks or soliton-like. This is indeed the case for the particles in the Ising field theory and the sine-Gordon model, which are solitons that interpolate between different vacua and are thus non-local relative to the field operators. In the case at hand, mesonic particles are not soliton-like particles and are, in this respect, local relative to the operator $S(x)$ (and other physical local operators that may be considered). Thus the matrix element $\langle n,p\vert S(x)\vert n,p\rangle_c$ should not have any singularities. This is also consistent with the absence of annihilation poles in the two meson electromagnetic form factor, see~\cite{einhorn1976confinement,jaffe1992when}~\footnote{Much like the 't~Hooft model, the attractive regime of the sine-Gordon model contains both non-local particles (solitons) and local ones (breathers, neutral soliton-antisoliton pairs), relative to the fundamental bosonic field in the Lagrangian, or local functions thereof, such as exponential vertex operators. When considering a local operator, such as the vertex operator, two particle form factors display annihilation poles when both particles are solitons, but not when they are breathers. This is a direct manifestation of the effect of locality, as discussed in the text: the solitons are non-local kinks with respect to the vertex operator, while the breathers are local. It is also worth noting that the vertex operator has a non-zero vacuum expectation value, much like our local observable~\eqref{vev}. See refs.~\cite{babujian1999exact,cubero2017quantum} for further discussions.}.

We can now rule out $\mathcal{O}(1/N_c)$ corrections that grow at late times in the 't~Hooft model. This rules out the scenario for thermalization proposed in eq.~\eqref{expthermalization}.

\subsubsection{Higher order corrections}

Of course, one can still argue that at some higher order in $1/N_c$ that a ``non-perturbative'' correction of the form~\eqref{expthermalization} will arise. With an understanding of the mechanism under which such terms grow at late times, we can schematically understand these higher order corrections. We need only to count the number of integrals over momenta and the annihilation poles of the matrix element. The leading contributions at late times will be those arising from matrix elements with the highest number of annihilation poles.   

Let us consider some examples.  Working at order $1/N_c^2$, the late time leading correction would contain connected matrix elements with two particles in both the incoming and outgoing states. These matrix element contain a single annihilation pole that can contribute to growth at late times. At order $1/N_c^3$ the leading term contains three particles in both incoming and outgoing states, and two annihilation poles arise. And so forth. As has already been argued in the Ising field theory and the sine-Gordon model~\cite{schuricht2012dynamics,calabrese2011quantum,calabrese2012quantum,calabrese2012quantum2,bertini2014quantum,cubero2017quantum,horvath2018overlap}, each new annihilation poles contributes a term that grows with higher powers at late times. It is then natural to expect contributions to~\eqref{perturbative} at late times are of the form
\begin{equation}
\langle S(x)\rangle_t=\langle 0\vert S(x)\vert 0\rangle+\sum_{i=1}^k\frac{\langle n_i,0\vert S(x)\vert n_i,0\rangle_c}{\langle n_i,0\vert n_i,0\rangle}+\frac{1}{N_c}A_1+\frac{t}{N_c^2}A_2+\frac{t^2}{N_c^3}A_3+\dots.
\end{equation}
Thus it is conceivable that the growing terms re-exponentiate at late times, as discussed in~\eqref{expthermalization}. However, as the $\mathcal{O}(1/N_c)$ term does not grow in time, the growth in time is shifted by one power of $1/N_c$ in the 't~Hooft model (as compared to Ising and sine-Gordon) and the non-perturbative correction arising from the exponentiation will affect \textit{only the $\mathcal{O}(1/N_c)$ term.}

We see, therefore, that the $\mathcal{O}(1)$ difference between the long-time expectation value and the thermal result cannot be overcome by the non-perturbative corrections, which modify only the $1/N_c$ correction, even at late times.  The expression for the expectation value~\eqref{perturbative} therefore holds true, even in the $t\to\infty$ limit. We can thus say that the thermal expectation value, \eqref{thermalexpectation}, is generally not recovered at late times after the quantum quench. The 't~Hooft model in the large-$N_c$ expansion therefore does not thermalize.

\subsection{Mesons as non-thermal states}
\label{sec:nonthermal}

In the preceding section, we have understood the reason behind the lack of thermalization in the large-$N_c$ limit of (1+1)-d QCD. Our central result, that states containing heavy mesons do not thermalize in the long-time limit, directly implies a violation of the (strong) eigenstate thermalization hypothesis~\cite{deutsch1991quantum,srednicki1994chaos,reimann2015eigenstate,deutsch2018eigenstate}: there must exist eigenstates within the spectrum where expectation values do not agree with the thermal average. This fits within a growing body of evidence~\cite{craps2014gravitational,craps2015holographic,dasilva2016holographic,myers2017holographic,kormos2016realtime,rakovszky2016hamiltonian,hodsagi2018quench,james2019nonthermal,robinson2019signatures,buyens2014matrix,buyens2016confinement,buyens2017finiterepresentation,buyens2017realtime,banuls2017density,akhtar2018symmetry,park2019glassy,mazza2018suppression,lerose2019quasilocalized} that system that exhibit confinement can exhibit an absence of thermalization, including some explicit examples that show violation of eigenstate thermalization~\cite{james2019nonthermal,robinson2019signatures}.

In the language of the eigenstate thermalization hypothesis, the result from the previous sections can be reformulated as the fact that there exist non-thermal eigenstates, $\vert \psi_{\text{NT}}\rangle$ of the full Hamiltonian, $H$ ,  of energy $E$, such that expectation values of local physical observables on this state do not agree with the thermal expectation value,
\begin{align}
\langle \psi_{\text{NT}}\vert O\vert \psi_{\text{NT}}\rangle\neq \langle O\rangle_\beta,
\end{align}
chosen to have matching average energy $\langle H\rangle_\beta=E$. These states, $\vert\psi_{\text{NT}}\rangle$, can be generated from the heavy-meson states of the previous section,  as
\begin{align}
\vert \psi_{\text{NT}}\rangle=\lim_{\tau\to\infty}e^{-\tau H}\vert n_1,0;n_2,0;\dots;n_k,0\rangle.
\end{align}
The process of taking the infinite-imaginary-time evolution of the heavy-meson states, amounts to selecting only the lowest lying eigenstate of $H$ which has an overlap with the meson state. From the results of the previous section we can conclude that
\begin{align}
\vert \psi_{\text{NT}}\rangle=\vert n_1,0;n_2,0;\dots;n_k,0\rangle+\mathcal{O}\left(\frac{1}{\sqrt{N_c}}\right),
\end{align}
that is, we have shown that eigenstates of the full Hamiltonian are well-approximated by the multi-meson states, being only weakly dressed by the interactions through $\mathcal{O}(1/\sqrt{N_c})$ corrections. This is contrary to what is expected to happen in theories satisfying the eigenstate thermalization hypothesis, where any non-thermal eigenstates of the free Hamiltonian, $H_0$ should receive non-perturbative $\mathcal{O}(1)$ corrections once integrability-breaking interactions are included, rendering the eigenstates of the full Hamiltonian thermal.

The idea that theories with confinement and emergent mesonic excitations do not obey the eigenstate thermalization hypothesis is (in hindsight) not so surprising. Consider the limit case of a single, large meson, there is a priori no reason to expect that expectation values within such a state should agree with a thermal average at the same energy density, where many excitations exist in thermal equilibrium. An important aspect for such states to remain well-separated from the thermal continuum is that the decay of single, high-energy meson excitations should be suppressed. In the case considered here, this is achieved under a large $N_c$ prescription, whilst in the now well-studied case of the Ising model~\cite{kormos2016realtime,rakovszky2016hamiltonian,hodsagi2018quench,james2019nonthermal,robinson2019signatures}, this can easily be achieved with a weak longitudinal magnetic field.   

There are numerous interesting directions to explore to further understand when, and why, mesons can be non-thermal. Tracing how the physics evolves away from the large $N_c$ limit would be interesting, although this requires non-perturbative methodologies. Hamiltonian truncation~\cite{james2018nonperturbative} may provide one such route, having already been applied to obtain low-energy  eigenstates of large-$N_c$ QCD~\cite{katz2014solution} and the Nambu-Jonas-Lasino model, which also exhibits meson and baryon excitations~\cite{azaria2016particle}. One might expect that at intermediate values of $N_c$ and at short-to-intermediate time scales, the physics still resembles the large-$N_c$ limit and statement made here apply to the prethermalization regime~\cite{berges2004prethermalization,moeckel2008interaction,essler2014quench,bertini2014prethermalization}. This would be worthwhile studying. Another interesting regime to consider would be the deconfined phase, where it has been known that heavy mesons can persist~\cite{asakawa2004jpsi}: Can these be interpreted in the framework of non-thermal states?

\section{Conclusions}
\label{sec:conclusions}

The large-$N_c$ limit of QCD has been well-studied since the seminal works of 't~Hooft in the mid-1970s~\cite{thooft1974twodimensional,thooft1974planar}. In light of recent works that have shown systems with confinement exhibit anomalous non-equilibrium dynamics and an absence of thermalization~\cite{craps2014gravitational,craps2015holographic,dasilva2016holographic,myers2017holographic,kormos2016realtime,rakovszky2016hamiltonian,hodsagi2018quench,james2019nonthermal,robinson2019signatures,buyens2014matrix,buyens2016confinement,buyens2017finiterepresentation,buyens2017realtime,banuls2017density,akhtar2018symmetry,park2019glassy,mazza2018suppression,lerose2019quasilocalized}, we have return to this theory to address the question of whether the 't~Hooft model thermalizes. Utilizing a powerful particle number/power of $N_c$ correspondence, and the annihilation pole structure of matrix elements within the theory, we have argued that the large-$N_c$ limit fails to thermalize. This statement holds even once non-perturbative corrections, which can arise at long times, are taken into account.

In particular, we have shown that simple initial states that contain few heavy meson excitations (with extensive total energy) do not thermalize. Here thermalize is used in the sense that expectation values of local observables do not relax, in the long time limit, to the relevant thermal result. At the heart of this physics is the local nature of meson excitations with respect to local observables, which implies there are no annihilation poles in matrix elements featuring one meson particle in both incoming and outgoing states. This absence of an annihilation pole prevents non-perturbative corrections that can bridge the $\mathcal{O}(1)$ gap between the $t=0$ initial state expectation value and the thermal result. Thus, for this class of initial states, thermalization is avoided. 

The absence of annihilation poles in one-particle-to-one-particle matrix elements is a crucial difference to analogous calculations in the Ising field theory and the sine-Gordon model~\cite{schuricht2012dynamics,calabrese2011quantum,calabrese2012quantum,calabrese2012quantum2,bertini2014quantum,cubero2017quantum,horvath2018overlap}. There the corresponding matrix elements contain singularities that arise from the non-local soliton-like nature of excitations, and these lead to non-perturbative corrections in the time-evolution at long times. This allows for a $\mathcal{O}(1)$ difference between initial and final expectation values in these systems. Our work thus provides insight into a general mechanism in non-integrable (1+1)-d systems with confinement whereby thermalization can be avoided.

These results lend further credence to the burgeoning literature on the absence of thermalization in models with confinement~\cite{craps2014gravitational,craps2015holographic,dasilva2016holographic,myers2017holographic,kormos2016realtime,rakovszky2016hamiltonian,hodsagi2018quench,james2019nonthermal,robinson2019signatures,buyens2014matrix,buyens2016confinement,buyens2017finiterepresentation,buyens2017realtime,banuls2017density,akhtar2018symmetry,park2019glassy,mazza2018suppression,lerose2019quasilocalized}. We do note that our results and arguments apply to the large-$N_c$ limit of (1+1)-d QCD, and their applicability to the more interesting case of (3+1)-d and particle-collider phenomenology is not so evident. In the context of the quark-gluon plasma generated in heavy ion collisions, for example, it is expected that the collision energy is high enough to enter the deconfined phase. There is no obvious reason to expect the class of initial states that we have studied, containing few very heavy weakly-confined mesons, to have any relevance in that case.  A more feasible future experimental testing ground could be quantum simulations of gauge theories with cold atomic gases~(see, e.g., refs.~\cite{zohar2013coldatom,surace2019lattice,celi2019emerging}), where engineering different non-equilibrium initial conditions and subsequent dynamics may be possible.

Additionally we have extended the concept of large-$N_c$ volume independence to non-equilibrium settings in confining gauge theories. These results are not limited to the low-dimensional 't~Hooft model, but can be applied to more realistic gauge theories in higher dimensions. This opens a door for future studies of QCD non-equilibrium dynamics, by restricting oneself to evaluating only the leading corrections in the $1/N_c$ expansion, corresponding to highly non-diagonal matrix elements of physical observables.

To finish, we note that non-equilibrium dynamics can be strongly affected by the presence of few, special states (which may, for example, be thermodynamically unimportant). The confining Ising model is a nice example of this: the family of single meson eigenstates that sit above the thermal multimeson continuum are unimportant in equilibrium probes, such as the dynamical spin-spin correlation functions measured in inelastic neutron scattering. They are simply not observed in such probes, being overwhelmed by the proximate-in-energy thermal states (see, e.g., Ref.~\cite{coldea2010quantum}). However, there is now plenty of evidence that such states can completely dominate non-equilibrium dynamics following simple quenches from simple, realisable initial states~\cite{kormos2016realtime,james2019nonthermal,mazza2018suppression,lerose2019quasilocalized,robinson2019signatures}. Thus when one finds interesting and unusual states, the question becomes: does my experiment realise and/or probe these?  While current experiments on the quark-gluon plasma likely do not probe the states suggested here, perhaps their presence may be observed in different experimental set ups or systems in the future.

\subsection*{Acknowledgments}

We are grateful to Andrew James and Robert Konik for useful discussions and comments on the manuscript. A.C.C. is supported by the European Union's Horizon 2020 research and innovation programme under the Marie Sk\l{}odowska-Curie grant agreement 75009. N.J.R. is supported by the European Union's Horizon 2020 research and innovation programme under the Marie Sk\l{}odowska-Curie grant agreement 745944.

\section*{References}
\bibliographystyle{iopart-num}
\bibliography{bib.bib}

\end{document}